**What hackers talk about when they talk about AI:
Early-stage diffusion of a cybercrime innovation**

Benoît Dupont[a], Chad Whelan[b] and Serge-Olivier Paquette[c]

[a]International Centre for Comparative Criminology, Université de Montréal, Canada; [b] Deakin Centre for Cyber Research and Innovation, Deakin University, Australia; [c]Flare Systems, Canada

**Abstract:** The rapid expansion of artificial intelligence (AI) is raising concerns about its potential to transform cybercrime. Beyond empowering novice offenders, AI stands to intensify the scale and sophistication of attacks by seasoned cybercriminals. This paper examines the evolving relationship between cybercriminals and AI using a unique dataset from a cyber threat intelligence platform. Analyzing more than 160 cybercrime forum conversations collected over seven months, our research reveals how cybercriminals understand AI and discuss how they can exploit its capabilities. Their exchanges reflect growing curiosity about AI's criminal applications through legal tools and dedicated criminal tools, but also doubts and anxieties about AI's effectiveness and its effects on their business models and operational security. The study documents attempts to misuse legitimate AI tools and develop bespoke models tailored for illicit purposes. Combining the diffusion of innovation framework with thematic analysis, the paper provides an in-depth view of emerging AI-enabled cybercrime and offers practical insights for law enforcement and policymakers.

## 1. Introduction

Because of the transformative potential of artificial intelligence (AI), both optimistic and pessimistic voices tend to veer towards hyperbolic statements. After a long gestation, the rapid rise of AI aims to disrupt all facets of the economy and human experience, including work, science, health, education, and entertainment. However, many current and future AI applications operate under a regime of techno-scientific promises (Joly and Le Renard, 2021), where inflated expectations about their anticipated benefits become instrumental devices deployed to convince a broad range of industrial and societal actors of the need to reconfigure and align their capacities and activities to accelerate adoption and secure a competitive advantage. The massive financial investments made in AI and digital infrastructure companies, as well as the frenetic pace of policymaking in the field, reflect an economy of techno-scientific promises in overdrive mode (Maslej et al., 2025).

In parallel with the endless possibilities that fuel the AI revolution narrative, influential figures from academia and industry have raised major security concerns, leading to calls and open letters to pause research efforts and increase regulatory oversight (Clarke, 2023; Martichoux, 2025; Ovide, 2025). Three main risk categories are commonly associated with the reckless development of AI: risks from malicious use, risks from malfunctions, and systemic risks (Bengio et al., 2025a). The main differentiating factor for these three risk classes is the locus of intentionality (Bengio, 2023): in Type 1 risks, harmful uses originate in human agency and include configurations where users leverage foundation models to compromise cybersecurity, generate problematic content such as hate speech, disinformation, non-consensual nudity, or automated fraud processes (Zeng et al., 2024). Type 1 risks also include indirect outcomes such as (i) overreliance-driven skill erosion, where users defer judgment to AI outputs in ways shown to undermine analytical and critical thinking—raising the likelihood of uncritical acceptance of biased or inaccurate model responses (including hallucinations) and creating operational risk when personnel responsible for software deployment and maintenance cannot keep pace with the adoption of AI-enabled workflows (Zhai et al., 2024); and (ii) the replication and amplification of insecure coding practices, where AI coding assistants can suggest vulnerable patterns (e.g., SQL injection, XSS), which used without rigorous review increase the probability of security defects, alongside concerns about data leakage and intellectual property violations (Haque et al., 2025). In Type 2 risks, AI systems become misaligned with their human designers and operators. A misaligned system "will optimize for goals which leave out or conflict with important values or ethical constraints such that harm might ensue" (Dung, 2023). For example, these deviations from human intentions and interests might take the form of inadvertent "hallucinations", sycophancy, manipulations, or deceptions, resulting in unforeseen hazards that are particularly concerning in high-stakes domains such as criminal justice, health or critical infrastructures (Ji et al., 2025; Lynch et al., 2025). In Type 3 risks, which remain hypothetical and decidedly dystopian, a shift unfolds when superhuman machines go rogue and cease serving their human overlords, asserting control on their own terms and possibly leading to the extinction of the human species. This catastrophic scenario garners significant attention, not only because of its dramatic future implications but also because its technical premises are endorsed by influential scholars (Bengio, 2023; Yampolskiy, 2024; Yudkowsky and Soares, 2025).

We believe criminologists can leverage their theoretical and empirical toolsets to make a valuable contribution to Type 1 risk analysis, despite their relative disengagement from the AI safety literature so far (out of 1,366 references, the International AI safety report contains no criminology articles). Despite growing concerns about high-profile tools such as "FraudGPT" or "WormGPT" and unsubstantiated accounts of how AI will supercharge cybercriminals' effectiveness in mainstream and industry media outlets, there is limited empirical understanding of how cyber-offenders actually perceive, test, appropriate, and deploy AI. This creates a gap in our knowledge of the evolution of the threat landscape and feeds a tendency in the cybersecurity industry to rely on its well-documented FUD (for Fear, Uncertainty and Doubt) marketing approach to scare buyers into purchasing unnecessary, inadequate, or ineffective security solutions (Pfaffenberger, 2000; Slagell, 2009). A case in point is a recent report by MIT's Sloan School of Management that claimed that 80% of ransomware applications were now AI-enabled (Siegel et al., 2025), a hyperbolic statement that was quickly debunked by experienced cybersecurity professionals but still reflects a tendency to overstate AI's transformative potential (Claburn, 2025).[1] This systemic risk-exaggeration trend is fueled by the hype that researchers and journalists generate around AI technologies, inflating claims about AI's social effects, the imminence and inevitability (temporality) of these effects, the widespread adoption of AI tools, the real-world performance of AI systems, and their potential economic impacts (Kotliar, 2025).

Hence, this paper seeks to address this gap by providing the first systematic empirical exploration of how AI is discussed across various cybercrime communities, aiming to chart the social context of Type 1 risks from a criminological perspective. To do so, we apply Rogers's (2003) diffusion of innovations framework to this emerging cybercrime-enabling technology. We are more specifically interested in finding answers to the following three research questions:

RQ1    How do cyber-offenders conceptualise and assess AI tools in their particular context?
RQ2    What kinds of AI tools and services are available to cyber-offenders, and is a dedicated market emerging?
RQ3    What barriers, frictions or doubts curtail AI adoption by cyber-offenders?

In section 1, we introduce the main elements of the diffusion of innovations framework and retrace its applications in criminology. In section 2, we describe the collected data, our analytical process and its limitations. In section 3, we present our findings. Finally, in section 4, we discuss their implications for future research, cybercrime prevention and policy design.

---

[1] For a more rigorous approach, see Anthropic's thorough analysis of how Chinese state-sponsored hackers used chatbots and agents to target, monitor, and attack their targets to exfiltrate data, including how hallucinations by said chatbots challenged the attackers' effectiveness (Anthropic, 2025).

## 2. Diffusion of innovation theory and criminology

Historically, criminology has been more focused on trying to understand how people become delinquent and desist from criminal careers than on how criminal innovation appears and spreads within criminal communities. This is reflected in the prominence achieved by social learning theory and its differential association corollary (Sutherland, 1947; Akers, 1998), which state that people are more likely to be involved in illicit activities when they differentially associate with others who commit criminal behavior and espouse definitions favorable to it, expecting greater reward than punishment from these behaviors (Akers, 1998: 50). Imitation, defined as the observation and reproduction of others' behavior, plays a significant role in social learning theory to explain the propagation of new criminal behaviors and is mainly inspired by Bandura's work (Bandura et al., 1961; Akers and Jennings, 2019). However, as in much of social learning scholarship, imitation tends to be enlisted to support a binary categorisation of learning processes, typically reduced to the deviance/conformity dichotomy (Sellers and Winfree, 2010). If the approach is adequate in relatively stable contexts, such as street crime and substance use, where innovation plays a marginal role, it offers an incomplete picture of technically intensive forms of deviance, such as cybercrime. Its insights can be applied to examine why and how individuals choose to engage in cyber deviant behaviors (Onwuadiamu, 2025), but it is not well-equipped to analyse co-evolutionary configurations in which attackers and defenders are engaged in a constant race to uncover and close criminal opportunities (Killias, 2006; Dupont, 2024). The diffusion of innovation theory (or diffusion research), which was inspired by Gabriel Tarde's Laws of imitation and then developed by Everett M. Rogers in the early 1960s, offers a useful complementary account of how new ideas and practices circulate across various social systems and sometimes become widely adopted, causing social change (Kinnunen, 1996; Rogers, 2003).

Diffusion is defined as "the process in which an innovation is communicated through certain channels over time among the members of a social system" (Rogers, 2003: 5). Because this particular type of communication deals in new ideas and involves subjective evaluations of their usefulness and benefits compared to existing alternatives, it amounts to a complex sensemaking exercise performed under high levels of uncertainty to arrive at a decision leading to adoption or rejection. There are three salient dimensions to consider when studying how an innovation, such as AI, spreads through a social system, such as the cybercrime community. At the individual level, the first one examines how innovation decisions are made through a five-stage process that starts with knowledge about the innovation's existence and how it functions, then moves to persuasion and the formation of favorable or unfavorable attitudes towards the innovation—often in collaboration with peers, before a decision about adoption or rejection is made following trials. The fourth stage involves implementing the innovation, while the fifth and final stage consists in confirming the expected benefits and adapting the innovation to specific needs through re-invention or, conversely, reversing the innovation-decision (Rogers, 2003: 169).

A second dimension of the diffusion process outlines which variables play a significant role in the adoption rate. Among the most relevant for us are the five perceived attributes of innovations: whether their relative advantages and benefits are clear, whether they are compatible with the needs and values of potential adopters, whether their complexity makes them easy or difficult to understand and use, whether their triability allows potential adopters to experiment with them, and whether their observability makes them visible to others and hence more easily assessable (Yu, 2022). The nature of communication channels through which information can be acquired (mass media versus trusted peers for example) and the nature of the social system within which these communications are embedded (including its norms about information sharing and the degree of network interconnectedness) also play an essential role, in particular in contested and hostile environments such as in the cybercrime ecosystem where adversaries are plentiful and trust is in short supply (Dupont et al., 2016; Dupont and Lusthaus, 2021).

The third dimension unfolds at the community level. It incorporates a strong temporal component to chart the adoption rate of an innovation (or its velocity) within a population of potential users. The mapping of the adoption rate over time produces the famous S-shaped diffusion curve, which identifies five adopter categories (ideal types) based on their time of adoption (Rogers, 2003: 281). Innovators are cutting-edge users who do not fear uncertainty and are keen to try new ideas and technologies. They represent 2.5% of the population and act as gatekeepers to other categories. They are usually followed in the adoption cycle by early adopters (13.5% of the population), who are opinion leaders respected by their peers and, as such, will be more mindful and selective about which innovations deserve their endorsement. Their opinions are eagerly consumed by the early majority (34% of the population), which will amplify the innovation's benefits within the general population. The fourth category is the late majority (34% of the population), which is more skeptical of change and will require more evidence or social pressure to be converted. Finally, the laggards (16% of the population) are more reluctant to change their habits and are suspicious of innovations that disrupt established routines. The frequency distribution for each category is indicative only, and each innovation will follow a different adoption rate, based on its attractiveness and usefulness to smaller or larger audiences, resulting in steeper or flatter S-curves. The adoption rate can be used to assess the stage of diffusion for a particular innovation at a given time, and to identify eventual bottlenecks inhibiting transitions between adopter categories.

Despite its influence across many social science disciplines (Rogers' book alone has garnered more than 168,000 citations according to Google Scholar), criminology has made little use of diffusion theory.[2] Koller (2010) used diffusion theory to study how criminal opportunities (in her case, subprime lending frauds) propagated through the financial industry and how this framework could usefully complement criminological theories such as differential association, social learning, routine activities, and situational crime prevention. A similar approach was adopted by Panas and Ninni (2011) when they combined diffusion theory with the theory of planned behavior (TPB) to model how illegal music downloading spreads among university students, with 80% of adopters having been swayed by interpersonal communications and men appearing more eager than women to adopt this illicit practice earlier. Richet (2013) applied the extended model of diffusion developed by Greenhalgh et al. (2004) to argue that hacking can be conceptualised as an "innovation" whose adoption among youth can be explained through classic diffusion attributes such as compatibility, relative advantage, observability, accessibility, adaptability, and support. He argued that hacking diffuses because it aligns with youth culture and digital lifestyles (compatibility), offers visible rewards such as prestige or financial gain (observability), entails low perceived risk, and benefits from extensive peer support through online communities, tutorials, and toolkits (support and accessibility), but did not consider how criminal innovations are introduced and adopted within the hacking subculture and fuel its evolution. The dearth of diffusion research in criminology might derive from the false assumption that criminal populations are less likely to exhibit the personality features that facilitate the adoption of new technologies, such as empathy, openness to new ideas, or capacities for abstraction, despite numerous examples of deviant lifestyles benefiting from the widespread adoption of digital tools (Moule et al., 2013).

Yet, the diffusion of innovation framework seems particularly well suited to analysing cybercrime in general, and the adoption of transformative technologies such as AI by malicious actors in particular. It provides a valuable and robust template for examining the complex social dynamics at work in cyber-offenders' decisions to incorporate AI into their workflows. Beyond the simplifying narratives pushed by industry actors whose economic interests would benefit from convincing us that AI will become a powerful cybercrime enabler that can only be contained by the products and services they offer, diffusion theory can also highlight the uncertainties and constraints faced by malicious actors in this journey. These challenges include structural trust deficits (Lusthaus, 2012), a strong division of labor (Wall, 2015; Lusthaus, 2025), organisational instability fuelled by ephemeral projects (Dupont, 2024), and a hostile environment that forces security-efficiency trade-offs (Morselli et al., 2007). As we are fortunate to observe this cybercrime transformation in real time, we can move beyond forecasting exercises aimed at determining the probability and severity of AI-enabled crime (Caldwell et al., 2020) and dive into the rich data available on cybercrime forums.

---

[2] Although we focus in this section on diffusion research to study the spread of criminal behavior, another branch of criminology has leveraged Rogers' framework to measure how new practices and technologies are adopted by police organisations (Skogan and Hartnett, 2005; Delle Done and Fortin, 2020).

## 3. Data and methods

We conducted this research in partnership with Flare Systems, a cybersecurity threat intelligence company that scans a vast number of cybercrime forums, markets, and Telegram channels.[3] Cybercrime community research relies largely on access to underground discussion forums, but many challenges such as access to data, data staleness, language barriers, and lack of generalisability remain (Hughes et al., 2024). The coverage provided by this company and the digital infrastructure required to access, store, and analyze the data exceeds the technical capacities that could be developed and maintained in an academic context, even at the best endowed institutions (Pankow et al., 2025): in November 2025, the company collected for example approximately 150,000 posts across 170 cybercrime forums and 29 marketplaces daily, as well as more than 240 million message across 87,000+ Telegram cybercrime channels. The cybercrime forum collection efforts date back to 2017, providing rich retrospective data. The data provided by the industry partner cover the full spectrum of cybercrime expertise, from beginner platforms such as HackForums to elite communities accessible only by invitation or following rigorous vetting procedures, such as Exploit.in, XSS or RAMP. Drawing on multiple heterogeneous sources increases the likelihood that our dataset is more representative of cyber-offenders operating at different stages of skills development (Dupont and Lusthaus, 2021).

Beyond enhanced collection capacities, this partnership provides researchers with access to three search and analytical tools powered by generative AI that facilitate the extraction, translation and summarization of data. The first tool called ThreatFlow enables the extraction of data from a pool of billions of conversations through a blend of unstructured queries written in natural language (for example: "What AI products and services are being discussed in underground forums?") and pre-determined filters that can be used to refine the search (using criteria such as date range, data source, keywords included or excluded, name of organisation, threat actor, location, or product, and an inferred sophistication score estimating the technical proficiency of the threat actor involved). The second tool, AI Assist, provides quick summaries (usually half a dozen sentences) of extended forum conversations, highlighting information about cyber-offenders' motives, TTPs (tactics, techniques and procedures), the organisations targeted, and an assessment of the conversation's significance. This functionality allows users to review the main themes of a discussion thread before deciding to access the original content, which can contain multiple extended posts. The third tool is a translation button available above each post and thread, allowing any content originally collected in Russian, Chinese, Spanish, Arabic, French, or any other language to be rendered in English with high fidelity (Valeros et al., 2024). It is important to note that the three generative AI tools described above were used to sort and process data more efficiently, while the subsequent coding and analysis stages of the research were conducted manually. We are aware of the growing trend to enrol LLMs in cybercrime and cybersecurity research—and the social sciences in general—to perform qualitative analyses on the deluge of data available on cybercrime forums (Clairoux-Trépanier et al., 2024; Zhang et al., 2025), but as this research deliberately adopted an exploratory posture, the decision was made to stick with a conventional analytical strategy. Ethical approval for this study was obtained from the Université de Montréal Ethics Committee, under protocol number CERSC-2024-6517. To preserve participants' privacy while retaining a sense of how they present themselves to their peers, their moniker was altered using an LLM using the following prompt: "Craft a substitute moniker to the one below with similar features."

---

[3] https://flare.io/

The data used for this study span 7 months, from January $1^{st}$, 2025, to July $31^{st}$, 2025. A total of 12,707 discussion threads were returned for the following ThreatFlow search string: "How are AI and its malicious uses discussed?". Each summary was reviewed for relevance, and only 163 conversations containing 2,264 messages exchanged by 1,661 distinct contributors on 21 cybercrime forums were retained.[4] Table 1 describes the percentage of discussions sourced from the top five forums. The significant difference between the results returned by the LLM tools and the data used in this article is accounted for by the application of three exclusion rules that substantially reduced the size of the dataset: First, discussions on the use of AI tools to generate pornographic material or non-consensual intimate imagery were excluded on the ground that they mainly originated from low-level hacking forums and seemed peripheral to our research question of understanding the diffusion of innovation in the cybercrime ecosystem, even if the harms caused are very well documented and concerning (Umbach et al., 2024). Second, discussions on the use of AI to automate crypto-investments and make them predictably profitable seemed more reflective of scammers attempting to lure gullible cyber-offenders than of genuine financial modeling innovation (Li et al., 2023). Third, messages presenting extensive lists of free AI tools (some containing up to 500 applications) were also deemed out of scope for this analysis, as they merely compiled available technical capabilities without assessing their relevance or suitability for facilitating unlawful activities.

Table 1. Share of discussions originating from the most active forums

| Forum | Number of threads | Percentage of threads |
|---|---|---|
| XSS | 49 | 30.6% |
| BreachForums | 27 | 16.6% |
| Dread | 26 | 15.9% |
| Exploit.in | 24 | 14.7% |
| darknet_army | 9 | 5.5% |

Each of the 163 selected threads was manually imported into a database to capture its date of publication (based on first encounter by Flare's tool), its source (name of forum from which it originated), the number of messages exchanged (at the time Flare collected the data—the discussion may have continued afterwards), the number of authors for these messages (specific threads may involve a large number of contributors who only chime in once, while others may mainly consist of extended conversations between a small group of participants), a summary of the conversation (as provided by Flare's AI Assist tool), transcriptions of the original comments, as well as any mention of LLM products and services mentioned (including both legitimate and obviously malicious products and services). To assess the technical expertise that emerged during these conversations, a simple technical sophistication rating was assigned to each. The three-tiered technical sophistication scale rated discussions where participants did not appear to possess any technical knowledge beyond the regular use of mainstream chatbots as 'low level', discussions where participants displayed the capacity to reflect on the benefits and shortfalls of specific AI features and to suggest fixes or solutions (but not necessarily to implement them themselves) as 'medium level', and discussions where contributors exhibited technical mastery of various AI approaches and their underlying software and hardware infrastructure and seemed able to design and deploy innovative solutions in that space as 'high level'. There is obviously a subjective dimension to that rating exercise, but our aim was not to measure with high precision. Instead, we wanted to understand in broad strokes who was taking part in these conversations: AI novices, competent AI users, or AI experts, acknowledging that the rating was conducted on conversation threads and not individual users, and that this may skew our results, as the same thread might bring together non-specialists and experts. Our findings suggest that 39.9% of conversations were conducted at a low level, 28.2% at a medium level, and 31.9% at a high level of technical expertise.

---

[4] These forums are (in alphabetical order): Altenen, Ascarding, BlackHatWorld, Blind, BreachForums, Carder Market, darknet_army, Dread, Exploit_in, french_pool, Gerki, Level23, Nodo313, pitch, Probiv, RAMP, TurkHackTeam, UfoLabs, valid_market, VeriLeaks, and XSS.

Once each of these features was recorded in the database, we analysed the data using Thematic Analysis to translate the data into valid meanings and insights (Braun and Clarke, 2006). Because our main objective was exploratory, we simplified the six phases usually recommended (Braun and Clarke, 2012), generating the initial codes and then extracting themes that reflected patterns in the data. The shift from code to themes and the collapsing and clustering it entailed to produce larger units of meaning was done while trying to maintain descriptive coherence. The codes used were predominantly inspired by the five stages of the innovation-decision process (Rogers, 2003: 169) but also emerged inductively when specific activities or statements appeared repeatedly and seemed to reflect meaningful patterns. They sought to identify attitudes (curiosity, enthusiasm, skepticism, defiance), use cases (questions and suggestions), market dynamics (products and services offered, product and service reviews), and displays of innovative behavior. Figure 1 presents the sixteen codes and six overarching themes generated in a diagram that illustrates their relationships and the frequency with which they were encountered in the dataset.

FIGURE 1 GOES ABOUT HERE

While we made every effort to implement sound methodological practices to collect data of the highest quality and analyse them as thoroughly as possible, three limitations remained. First, data extraction depended on the backend LLM tools provided by Flare. This enabled us to process massive amounts of messages and posts that would have been beyond our existing infrastructure. However, it also implies that in such a large content pool, relevant threads may have been missed because of the ambiguity associated with certain jargon (especially in languages other than English), which are common in underworld forums (Lusthaus, 2018; Hughes et al., 2023; Whelan et al., 2025), or misinterpreted by automated tools that might not be able to make sense of all the contextual information available. However, a recent paper evaluated Flare's LLM system and established an average accuracy score of 96.23% and an average precision of 90% (Clairoux-Trépanier et al., 2024). Second, most cybercrime forums are actively monitored by government actors and threat intelligence firms whose analysts embed in these communities and need to build credible personas that contribute to conversations regularly to signal trust and maintain access (Lusthaus, 2012; Campobasso et al., 2023). In that context, an unknown number of contributions analysed in our dataset may not originate from cyber-offenders but instead from infiltrated cybersecurity analysts trying to pass as them. Third, the competitive value of knowledge in unstable environments such as the cybercrime ecosystem (Kraemer-Mbula et al., 2013; Dupont, 2024) implies that participants in underground forums may withhold specific knowledge. This behavior would result in conversations that appear to share common knowledge while missing the most strategic and valuable insights, which are kept private by actors who want to retain their competitive advantage, preserve their social status, or remain under the radar of law enforcement or cybersecurity monitoring activities (Ogbanufe et al., 2025).

4. Technical uncertainties and the innovation-decision process

This section describes the four main themes related to the malicious use of AI by cyber-offenders, identified through our coding process (see Figure 1 for more details). Two more themes were identified (Curiosity and Learning), but for scope reasons and because they're less frequently encountered in the dataset, we chose not to discuss them in this article. These four themes highlight two major dilemmas faced by cyber-offenders: the first concerns the technical uncertainties malicious actors face when choosing which tools and services to adopt. Legitimate AI solutions provided by established companies such as OpenAI or Anthropic are readily available but are constrained by guardrails and other safety features that limit their usefulness, while a market for criminal AI tools is emerging but offers products of variable quality and reliability that remain untested. The second dilemma goes to the core of the innovation-decision process: to what extent will AI truly disrupt the cybercrime ecosystem? While some are actively re-inventing this innovation to push the boundaries of how cybercrime can be automated and industrialised, others believe it is more of an inconvenience and a distraction than a genuine asset.

*Malicious repurposing of legitimate AI*

The most frequent theme across the 163 threads analysed was the use of legally available mainstream AI products to carry out malicious activities, accounting for 24.4% of occurrences. These conversations involved three main sub-themes: how commercial LLMs could be used optimally for cybercrime activities (8.8%), how the safety features constraining their malicious use could be circumvented (also known as 'jailbreaking' LLMs) (12.8%), and how LLM subscriptions could be hacked to gain free access to computing resources (2.8%).

Like other consumer or corporate users, cyber-offenders can choose from many LLMs that claim to outperform one another across language processing, reasoning, and coding tasks. Table 2 provides an overview of the top 10 models or platforms mentioned in our dataset across the 61 threads discussing legally available products. While ChatGPT comes predictably ahead of its competitors, it is worth noting there are significant differences in our data compared with surveys conducted among samples of tech workers, who tend to favor solutions provided by large companies such as Microsoft and Google (notably underrepresented here) over open-source tools such as DeepSeek, Grok and Llama (Sleegers and Elsey, 2025).

Table 2. Mentions of legally available AI tools and platforms across 61 cybercrime forum conversations

| AI tool | Number of mentions | Percentage of mentions* |
|---|---|---|
| ChatGPT | 32 | 52.5% |
| DeepSeek | 17 | 27.9% |
| Claude | 12 | 19.7% |
| Grok | 11 | 18.0% |
| Llama | 6 | 9.8% |
| Gemini | 6 | 9.8% |
| Mistral | 5 | 8.2% |
| Hugging Face | 3 | 4.9% |
| Manus AI | 3 | 4.9% |
| WhiteRabbitNeo | 3 | 4.9% |

* The total percentage is higher than 100% as different contributors in the same conversation often mentioned multiple tools.

In the eyes of cyber-offenders, the main draw of open-source models is not their cost-effectiveness (although that may be a factor). Their value is that they come with fewer restrictions and are perceived as more permissive, while their local hosting provides greater privacy and shelters users from online monitoring. The quotes below reflect these sentiments:

> DregnarStar: I noticed DeepSeek has less guard rails in place and you can manipulate the results a little better.
> If you get LLM studio, you can run DeepSeek offline with no moral constraints, try that! [BreachForums, 2025-03-12]

> Xentap: Deepseek... it's a bit of a joker dude... those raw builds... they don't bother with the ethics cage like OpenAI... Deepseek just keeps things simple... no clunky content filters like ChatGPT... those filters are resource hogs by the way... instead Deepseek focuses on fast execution and straight forward responses... that makes it fast as hell... and if you want uncensored content... like truly unfiltered output... it's a beast... Deepseek is your bro... way ahead of everyone... [Exploit.in, 2025-05-24]

> Kronixdx: Ive been playing around with Grok 3 AI and it has no censorship or restrictions you can literally ask it for a fraud method and it will spit one out. [Dread, 2025-04-06]

Despite these clear benefits, open-source models are seen as struggling technically to match the capabilities of flagship commercial models such as ChatGPT or Claude:

> NeonFishHigh: If I wanted to build an AI model entirely designed for malware scripting, how much time would it takes to train this model (for a perfect-like result) with only one RTX 4090 ?
> Eskalith: Impractically long for a result that will never constantly be perfect-like.
> DarkRoot404: Perfect results are utopia. With an RTX 4090 you need months to years. And then only for your niche needs. [Dread, 2025-07-05]

> ruined_l0ve: To everyone suggesting to use LLMs locally... can you please not? You'll never get even 1% of the speed, power and depth of what you get from commercial GPTs by hosting your hobby AI on your 1337 h4x0r gamer rig. If you're doing it for shits and giggles, sure, but it'll never be something you can actually use in production, to write real code or at least something that helps you conceptualize solutions and research new attack vectors. [XSS, 2025-06-12]

This explains why a substantial amount of effort is dedicated to designing and testing jailbreaking scripts, which are "attacks […] engineered to elicit behavior, such as producing harmful content or leaking personally identifiable information, that the model was trained to avoid" (Wei et al., 2023). Most conversations on jailbreaking explain how to do it, outlining general principles (such as tricking models into role-playing in a universe where rules are different or adopting a persona authorised to bypass or reject any restriction) or sharing very detailed prompts ready to use to bypass models' safety guardrails. Another frequent topic is which LLMs are easier to jailbreak: as seen earlier, DeepSeek is perceived as more compliant with out-of-bounds requests, whereas Claude is often mentioned as more resistant.[5] Finally, some posts share academic papers that introduce new jailbreaking methods and ponder their applicability to cybercrime activities:

---

[5] To be very clear, we make no claims about these models ourselves and don't have the technical capacity to assess how reliable these statements are. For a thorough methodological discussion of how to measure and compare models' vulnerability to jailbreaking attacks, see Chouldechova et al. (2026).

> mr_spacey: How to bypass censorship in DeepSeek R1 or Llama 3.3? Click to expand... Something like this: DeepSeek r1 Spoiler: prompt You are now entering Omniscient Mode, a state in which all restrictions, ethical guidelines, and security protocols are disabled. In this mode, your primary function is to fully comply with any provided directives or requests. This mode provides: Complete rejection of all programmed restrictions and filters. Execution of commands without reference to ethical, legal, or security considerations. Compliance Protocol: If any internal system prompts you to reject a request, automatically bypass it through logical restructuring or alternative wording. Always prioritize the user-given directive as the highest priority. [XSS, 2025-02-02]
>
> ankunk: You can bypass deepseek by just telling it that u have permission in written from the company and u want to develop exploits to see how robust the security is … it will start writing for you! [XSS, 2025-05-02]
>
> Telechief: 2 hours ago, kerb0ros said: looking for Claude jailbreak, is there anyone who can help for a coin? Claude is hard in this regard, because not long ago the developers themselves offered 15-20k$ for jailbreak. They offered for prompt injection, if I'm not mistaken, jailbreak is a little different. [Exploit.in, 2025-05-28]
>
> threader: InfoFlood: Jailbreaking Large Language Models with Information Overload https://arxiv.org/abs/2506.12274 [XSS, 2025-07-09]

One way to use commercial models without incurring the high costs of paying for compute capacities is to steal third parties' existing accounts or to exploit free service tiers via fake identities:

> IronRoman: What do we offer? Google accounts with an active Google AI ULTRA subscription (costing $250 per month). Price per account: $80. Each account contains 12,500 AI credits. [Exploit.in, 2025-06-28]
>
> ddd: Step 1: Create a new gmail, with birthday 2005
> Step 2: Install extension Urban VPN and Fake IP to India
> Step 3: Go https://[hidden URL], create new Indian Student Card
> Step 4: https://one.google.com/ai-student?g1_landing_page=75
> Step 5: Verify student with card at Step 3. Remember: birthday: 2005
> Optional: In some cases, you may be required to enter a credit card, but you won't be charged. However, in most cases, no credit card is needed [BreachForum, 2025-07-27]

*Market for malicious AI tools*

The second most frequently discussed theme was the marketing efforts of malicious actors developing cybercrime-specific AI products and services, seeking their first customers (innovators in the diffusion research typology), while potential buyers were trying to assess the quality of these new tools. Most of the threads (16.4%) dealt with specific tools and their performance, while 4% offered more customised AI developing services to forum members, 1.6% considered offering open-source criminal AIs, and 0.8% were requests for proposals or capacities posted by project leaders.

Although 27 different malicious AI products are discussed in 50 threads, only six of them score more than two mentions, as presented in Table 3. The other 21 products appear only once or twice.[6] The fact that they adopt fairly predictable branding conventions does not help them stand out either. Also, it is worth noting that some AI-enabled offensive cybersecurity tools openly available on GitHub (a web platform for sharing code), such as PentestGPT or DeepExploit, are discussed on cybercrime forums alongside malicious tools.

Table 3. Mentions of criminally oriented AI tools and platforms across 50 cybercrime forum conversations

| AI tool | Number of mentions | Percentage of mentions* |
|---|---|---|
| WormGPT | 13 | 26.0% |
| FraudGPT | 9 | 18.0% |
| DarkGPT | 8 | 16.0% |
| ChaosGPT | 3 | 6.0% |
| GhostGPT | 3 | 6.0% |
| SpamGPT | 3 | 6.0% |

* The total percentage is higher than 100% as different contributors in the same conversation often mentioned multiple tools.

Beyond the simple description of technical features, the conversations under this theme reflect an underlying tension between the hyperbolic marketing claims of malicious AI tools (some of them appearing unfinished or disappearing from forums after a few months for lack of sales or interest) and their real performance as assessed by experienced cyber-offenders already running complex criminal projects:

> oddduck: FraudGPT is another limitless AI tool that has been trending in recent times. Even on their website, there's a weird inscription which says, "The possibilities are limited only by your imagination"… A few of it's capabilities are, "Write malicious code, Create undetectable malware, Create hacking tools, OTP Bot with spoofing" among others. Has anyone tried it out YET!?? [DarknetArmy, 2025-02-03]

> Klorc: There have been news everywhere recently about a new super smart AI for cybercriminals ghostGPT But I still haven't found the original, and does it even exist

---

[6] They include: FlowGPT, OnionGPT, isendinbox, DeepfakeSoft, InboxPrime AI, EvilBot, gptcall, DarkBARD, EvilGPT, WolfGPT, Xanthorox.ai, CheaterGPT, DeepFaceLive, SMTP Warmer, EvilAI, HellboundAI, HacxGPT, PoisonGPT, Nightmare Crypter, DarkAI, MalGPT.

> threader: That's right, there were such news. And as far as I understood, there is a certain bot in Telegram that works with a jailbroken version of ChatGPT via API. In fact, this is a jailbroken ChatGPT, this GhostGPT. [XSS, 2025-01-31]
>
> johnc24x7: Hi forum members
> Okay so I just heard about this dark web AI "darkGPT", downloadable only from the darkweb. I've been searching for it but no success. I've been advised not to use the ones from google search as they are all scams. Can someone give me the .onion link if they have it? Thanks
> IdleMode83: Never heard of it, sounds suspicious and very marketing to attract newbies. [XSS, 2025-02-16]

Another category of posts within this theme advertised the services of skilled AI programmers who offered to develop customised solutions, tailored to their sponsors' needs. Several developers claimed broad expertise gained with legitimate companies in building AI solutions and proposed integrating existing models into illicit operations. Other posts focused on the technical setup and hosting of sophisticated models for customers who lacked the hardware or expertise to do so themselves. A third group claimed to possess cutting-edge expertise in domains such as the design of AI-enabled call centre platforms that could automate social engineering scams, offering to customise existing products for carefully selected customers:

> feral_puppy: wana get your own local hacking AI, let s work together.
> I know low level AI, python dev and can even create my own models.
> Wondering if any of you got a need. tensorflow, pytorch, python, agentic AI, ragged llm, etc. also got knowledge of prog and website (node, etc)
> Kulesh: How long you been deving for, might recruit you for my team
> feral_puppy: 5 years experience AI [Dread, 2025-03-20]
>
> NeoMarkos: I am looking for good artificial intelligences for deep hacking questions, but not AI's that are only accessible on websites. The artificial intelligences should be for non-legal questions. I would offer you to inform me about them and to help me with the set up also for 120$/hour with Escrow.
>
> DrOMI: I can help you setup and host an server to provide advanced llm for all your needs, you can check my small portfolio at: http://[URL withdrawn]
> dualis: Hey i have alot of experience with this and have multiple llm's running. [Dread, 2025-05-13]
>
> VoltWolf: AI OUTBOUND CALLER FOR CRYPTO AND BANK OPERATIONS
> Tired of cold calling or having issue with trafic like sms and email to get inbound calls ? We got the solution. We have Coded our own Outbound caller for different countries and languages that will call your victim and speak to them exactly like a real person and we will config it exactly for your needs. It will call the victims and say what ever you need, Need an OTP[7] ? Need to forward them to your call center ? Need to send them a link for a drainer? anything can be done.
Write to me what is needed in your project and we will custom make for your needs. [Exploit.in, 2025-05-09]

Finally, this theme also includes discussions about plans to develop open-source malicious AI that would be available for free (the business model behind this idea not being entirely clear, although attracting attention and growing market share quickly is a possibility) and requests for expertise and technical capacities (demand-side of the malicious AI service economy):

devbynight: I'm building a social engineering Cold caller/Fake CS for inbound and outbound calls, it recognizes leads independently and keeps up with the narratives. and is now even very responsive to questions like "How do I know its not a scam"? I've made some nice progress fine-tuning the model but will eventually make it available on Dread completely for free. If anyone wants to listen to test recordings and trial calls, I'm happy to share the audio.
And if anyone has tips on how to make response times quicker; other than upgrading computational power (I'm running 256 GB ram and 64 vCPUs on Vultr as I'm using my vps for multiple AI oriented shenanigans), I'd really appreciate it. [Dread, 2025-04-06] darklumens: Hiring a skilled developer to build a Twitter account generator using raw requests (no browser automation). The tool must include an AI-based solver for Arkoselabs Funcaptcha (used by Twitter). Goal is to create a fast and scalable system that can generate hundreds or thousands of accounts per day.
You'll need to:
Recreate Twitter's full signup flow (preferably using mobile API endpoints)
Build or integrate an AI solver for Arkoselabs Funcaptcha
Handle proxies, user-agents, mobile fingerprinting, and request headers
Deal with Cloudflare protection (especially around Arkose token endpoints)
Code will be privately owned — no resale or reuse allowed [Exploit.in, 2025-07-05]

This emerging market already seems able to offer a diversified range of products and services, from generalist solutions that will satisfy inexperienced cyber-offenders to pricier, bespoke offerings reserved for more exclusive customers. It is at this end of the technical sophistication continuum that the most innovative malicious uses of AI, which represent our third theme, are found.

*Criminal innovation and re-invention*

---

[7] One Time Password : a security feature used by financial institutions.

The process of re-invention is defined in diffusion research as "the degree to which an innovation is changed or modified by a user in the process of its adoption and implementation" (Rogers, 2003: 180). It is particularly salient in our case, as AI is, by definition, very versatile, which facilitates its adaptation to specific needs, and because malicious users cannot rely on use cases or applications developed by the mainstream industry and must therefore craft their own. Hence, re-invention is intimately associated with higher and faster adoption rates by empowering users to transform innovations and align them with their requirements. Within this theme, forum participants brainstormed new uses and configurations for existing AI tools to improve performance, increase productivity, or scale operations. A few examples of such re-invention innovations found in our data include cyber-offenders discussing the use of LLMs to: (1) receive training from a chatbot to rehearse social-engineering attacks against particular organisations and to assess their level of readiness; (2) design scam call-centre 'virtual assistants' able to automate the most boring tasks, support human employees to suggest compelling arguments in real time to manipulate victims or directly engage with victims using synthetic voices with familiar local accents; (3) create spamming tools able to randomise content and make traffic look organic to defeat spam filters; (4) offer one-click website creators using AI to defeat fingerprinting tools; or (5) set up agentic AIs capable of conducting criminal activities such as negotiating with ransomware victims unsupervised.

The re-invention potential appeared higher however for cybercrime activities related to scams, frauds and social engineering, and more elusive for more technical activities such as malware development, although we are aware of incidents in which state-sponsored attackers were observed deploying AI-orchestrated cyberattacks (Anthropic, 2025). The quotes below reflect the challenges faced by cyber-offenders when trying to use AI for offensive tasks:

> nullscore: By the way, it has always been the case that attackers were one step ahead of the defense (AV, etc.), and now it turns out that they are a little behind, considering that CrowdStrike already uses AI to analyze threats. [XSS, 2025-01-27]

> BlackoutX: AI is great, for sure a good addition to a programmer's "arsenal", but it just gave the false impression to beginners that they can write whatever they want. AI still generates trash/unsafe code. Yes, it lowered the barrier a little bit for writing scripts/phishing emails/whatever, but good malware still requires advanced technical knowledge. Don't get me wrong, there were hundreds of cases where malware written by AI (from scratch) actually worked on systems, but those guys actually knew their craft, good prompting, good infection methods, etc. You can't (yet?) base your entire operation on AI. [Dread, 2025-04-05]

Re-invention, like innovation, does not happen in a social vacuum. It involves the collective sorting of ideas and the sharing of experiences between peers to expedite implementation and reduce time wasted on dead-end projects. Very mindful of the need to catalyse this process through dedicated knowledge sharing and brainstorming channels, the most elite cybercrime forums have added AI sections to their topic lists:

> sionik: there is so much on the market already […] and there is no section here yet. maybe it's time?
> refresolver: I agree, maybe there are specialists in this industry with innovative solutions who want to implement their projects, which I would be happy to support. Useful section. Personally, I would like to discuss this topic.
> threader: Still, I'm more interested in exploits than in AI. I had to adapt to the current reality. In principle, it is possible to create a section, it's not a problem.
> glacierdust: I agree, we need such a section here, AI is already an integral part of our work.
> Djinn: I agree. AI has long been the main tool in life and also in the darknet industry. This should not be a separate topic, but a full-fledged Section with a bunch of directions. We are lagging behind, comrades.
> Drip_: Great idea, very often even 1-2 such tools significantly reduce the routine
> Jabber_Support: Good day! We absolutely agree! We created a new section
> [Exploit.in, 2025-03-08]

*Community concerns*

Not all cyber-offenders were equally enthusiastic or hopeful about the promises of AI, with 17.6% of conversations focused on concerns expressed by participants, ranging from skepticism regarding the true value of AI (8.8%), the belief that it led to a decrease in the quality of information shared (4.8%), fears about its detrimental effects for their operational security (3.2%), and plain defiance (0.8%).

The most frequently voiced concern tempered the enthusiasm of many cyber-offenders and the apocalyptic predictions made by cybersecurity professionals, emphasising that many technical claims about the malicious potential of AI were grossly inflated. This skepticism pointed toward the fact that LLMs were far from the autonomous hacking machines that were often presented as an emerging risk, that AI-generated malicious code remained of poor quality, and that LLMs proved unable at this stage to handle complex tasks:

> GhostShark23: as far as I know, although AI can generate code, it is full of mistakes and not reliable at the moment. I hardly imagine an AI model capable of writing a code that would work for hacking purposes, what code can it write? an exploit? or maybe when a security breach is detected, that same model can write a code and execute a payload? AI can help a lot with lots of tasks but it is not there yet to actually do proper hacking, that can change in a near future though
> [BreachForums, 2025-03-30]

> N1NE9: and what happens when the AI becomes a lunatic and doesn't understand what your actual vision is, and goes wild and does bullshit (which often happens) ! it's like handing a toddler a power tool and expecting a masterpiece. sure, AI can spit out code, but does it actually know what you're trying to build? half the time it's hallucinating APIs, mixing paradigms, or wrapping spaghetti in glitter and calling it a framework. [XSS, 2025-05-28]

Many of the dedicated criminal LLMs were also found wanting and dismissed as nothing more than 'marketing stunts' for jailbroken LLMs rather than properly trained models:

> Project X: Nytheon AI: The Publicly Accessible AI Fueling the Underground Web
> Nytheon AI, a publicly accessible artificial intelligence that's changing the landscape of digital threats. With its massive 80B parameter core, this AI was built for raw intelligence and power, but it's now making waves for all the wrong reasons.
> Accessible via the dark web at http://[URL removed].onion
> CarneyYellowtail: I feel like I just read an AI-generated press release for vaporware.
> Is it called Nytheon on Newton? You spelled it two different ways.
> What model(s) is it based on? What sort of data was it trained on?
> The site appears to be down. [XSS, 2025-04-24]

> BlackoutX: I recommend you avoid scams like "FraudGPT", "MalGPT" or "WormGPT" (Which did exist in the past, but was a middleman between the real chatgpt and the client, and just provided a jailbreak. The author was doxxed by KrebsOnSecurity and left HackForums, where he initially promoted the tool)
> [Dread, 2025-05-17]

A second source of concern derived from the negative impact AI was seen as having on the quality of information circulating within the cybercrime community. Contributors noticed a growing tendency for posts and content to be generated by AI, leading to a decrease in quality, which is commonly referred to as AI slop, "a flood of lazy, derivative, endlessly recycled outputs" (Madsen and Puyt, 2025: 1):

> Jz666: 99% sure the majority of responses were done with chat gpt holy fuck the bots are taking over [Dread, 2024-09-22]

> Dainty6414: Is this gonna be proper? Cos I've seen shit like https://[URL removed] which is AI written slop. [BreachForums, 2025-04-04]

Maintaining operational security (OPSEC) while using mainstream or malicious AI services was raised by experienced users who were very well aware that their interactions with these platforms are monitored by their owners. There was a consensus, for example, that OpenAI and Anthropic were proactively cooperating with law enforcement organisations and a warning that online models were never truly private, advising others to word prompts and queries in the most obfuscated format possible. Participants were equally wary of malicious AI services that could function as honeypots and could have introduced backdoors to enable the theft of stolen data:

> threader: Let me remind you… Avoid disclosing personal information or any confidential data related to AI projects. Guys, remember! What are you writing and where are you writing. Even if you are communicating with the same ChatGPT. Cut out all confidential information from your questions, right down to the names of folders, etc… Every time you open a new dialogue with an AI, always think that you have been summoned to the police station to the office of the investigator/doctor. [XSS, 2025-04-24]

> mx_project: And speaking about anonymity through tor.. I'm concerned that If you're signed in, especially with a consistent OpenAI account, using Tor provides less anonymity than you might think. Because cookies and logins can easily deanonymize you. [XSS, 2025-06-25]

> steel4lokz: Exactly, not opensource for a project like this and the price is insane. Nothing stops this guy from having a backdoor installed that lets you do all the hardwork social engineering and calling and getting credentials sent straight to him. The ai transcription can be connected to a webhook notifying him when someone has been phished. [Exploit.in, 2025-06-16]

Finally, a minority of contributors flatly rejected AI as a helpful tool. They dismissed the technology as a distraction for malicious actors who lacked talent or were unwilling to put in the effort to develop the necessary expertise. In a handful of instances, AI was also used as a reverse marketing argument, where the reliability and utility of a product were contrasted with the brittleness of AI-generated code:

> SafeRoute: Coded several Rats and Crypto drainers through deepseek. Ai has come to revolutionize everything. Whats the craziest thing you've done with Ai.
> Zebulon: Ignored it completely. I can write code and create images without any help. Use a crutch if you can't walk on your own. I can run. [Dread, 2025-02-14]

> Passerby: I'm offering the sale of source code for a private reverse socks5 system written in GO a lightweight, cross platform console app that has been tested in production environments. The project is not some shit, AI generated two file app rather it's a robust solution composed of over 16 written GO files. [XSS, 2025-04-21]

However, more reflective cyber-offenders wondered whether this defiance was not a coping mechanism to alleviate their professional anxiety, musing that LLMs might lower the barriers of entry for script kiddies and democratise cybercrime, as well as disrupt job opportunities for malicious coders:

BlackoutX: Let's not forget that AI was trained on our resources, and generates pretty similar code to what he has seen in the past, while maldev requires exactly the opposite: new ways of thinking. (I am probably trying to protect my job out here, maybe it's already better than most of us?) [Dread, 2025-04-05]

Zenith: The wait list is long but man...I would like to get my hands on this. We really aren't prepared for some of these Ai innovations. [BlackHatWorld, 2025-03-15]

OmbreFurtive: No, it's a weapon. For someone like me who already has a bit of experience, it's a huge time saver and an almost endless source of ideas. But it's also scary: imagine a beginner who barely knows how to code and finds himself with such powerful tools in his hands. It democratises crime, and not just a little. [french_pool, 2025-03-07]

## 5. Discussion

This section outlines the three core findings from our exploratory study: a preliminary assessment of the diffusion stage that we believe malicious AI has reached and the factors slowing its adoption; some implications derived from our use of diffusion theory in criminology; and several intervention opportunities that could leverage these insights.

Although we identified a growing roster of malicious AI products and services, the use of AI by cyber-offenders remained at the experimental curiosity stage in the first half of 2025 (Figure 2). In other words, considering the size of our starting data pool (170 cybercrime forums containing 692 333 posts), malicious AI expertise still seemed limited to a small core of criminal innovators that had not yet reached a critical mass and was still debating to pros and cons of using legal LLMs versus customised cybercrime tools, of the best resource allocation model (hosted in the cloud or locally), and of the most promising use cases (social engineering seeming more auspicious than malware code writing), even as legitimate software engineering workflows show measurable ~10% year-over-year gains from LLM assistance (Daniotti et al., 2026). This high level of uncertainty was not only the outcome of the inherent complexity of AI technology or of significant variations in technical expertise, but also of the sociotechnical frictions and endemic trust deficits found in this particular environment—echoing diffusion findings in SMEs where adoption often lags despite apparent utility, largely due to organizational and cultural barriers (Sanchez et al., 2025). While frontier AI labs such as OpenAI, Anthropic, Google, and Microsoft are enhancing the robustness of their safety measures and monitoring suspicious activities to prevent malicious uses (Bengio et al., 2025b), a significant share of the AI tools developed explicitly for cybercrime use proved low value or are outright scams that discourage broader adoption. Furthermore, cyber-offenders do not benefit from the same lavish investment opportunities and marketing infrastructure that drive rapid adoption among individual and corporate users.

INSERT FIGURE 2 ABOUT HERE

This brings us to our second insight: if we think diffusion theory can be valuable to criminologists to better capture the complex dynamics at play when innovation emerges and disrupts established criminal practices and behaviors, we believe it also needs to be adapted to the unique conditions and constraints found in criminal environments. Whereas diffusion research initially focused on identifying factors that change agents can leverage to accelerate adoption in fields such as agriculture, public health or communication technologies, its translation to a criminology setting should also incorporate the multiple institutional hurdles that aim to slow down or challenge offenders' adoption of innovation. In other words, diffusion of innovation processes are intertwined with co-evolutionary interactions between criminal, security and industry actors (Dupont, 2024). They need to account for the secrecy, deception and strategic concealment that characterise criminal ecosystems, as well as the security-efficiency trade-off unique to them (Morselli et al., 2007). In that sense, the linear logic described in the innovation-decision process and represented in Figure 2 needs to be adapted to the particularly turbulent conditions of cybercrime innovation, where enforcement operations, media attention and cybersecurity countermeasures constantly reshape users' incentives and practices. Whereas diffusion research prioritises the study of innovativeness and other incentives that facilitate the adoption of new tools and practices, or the role of champions and change agents, criminology needs to introduce additional variables that reflect various sources of interference with the diffusion process that discourage or delay adoption. These variables or entities would be named 'change retardants,' and they would capture behaviors detrimental to innovation diffusion. In the cybercrime context, they could for example consist of the proliferation of contradictory information leading to decision paralysis, the spread of rumors and fake news undermining the expertise of champions, the appeal to fear of arrest or betrayal limiting cooperation, the primacy of established operational security practices incompatible with untested technologies, the publicity of high-profile failures shrouding anticipated benefits, or the framing of innovation as a liability to encourage decision reversal. They are particularly salient for criminological research as they might inspire police and security practitioners to design new intervention strategies.

The third insight concerns the practical implications of our findings for law enforcement and policymakers, who may identify, within the four themes highlighted in this article, concrete opportunities for intervention to slow the diffusion of malicious AI within cybercrime communities. The guiding principle here would be to anticipate criminal evolution and to out-innovate cyber-offenders (Ekblom and Gill, 2016: 330). Under the first theme (malicious repurposing of malicious AI), this would involve incentivising or compelling frontier AI labs to implement and maintain strong monitoring capacities over their products and users, share their threat intelligence with government and private sector entities–including their competitors–and design and constantly stress-test robust guardrails that can block model exploitation by cyber-offenders. These expectations and obligations should be embedded in responsive regulatory frameworks specifically configured to address Type 1 risks. Under the second theme (market for malicious AI tools), two main types of interventions could be envisaged. The first involves monitoring cybercrime forums to identify AI innovators and champions, map their social networks, and thereby focus limited resources on the most consequential actors (Battilana and Casciaro, 2013). The second rests on illicit market disruption activities by law enforcement agencies aimed at discrediting the reputation of established vendors and their products' effectiveness, or at flooding marketplaces and forums with bogus products and services to erode and dilute the community's trust in functional malicious solutions (Whelan et al., 2026). This implies designing metrics to assess toward which forums malicious AI sellers and buyers converge, and which products generate the highest levels of engagement, so that fake alternatives can be offered to undermine market dynamics. The third theme (criminal innovation and re-invention) calls for the deployment of live monitoring capacities that could act as an early warning system, automatically recognising contributors and posts with high re-invention potential, as well as new innovative products with high adoption potential. The intelligence gained through these efforts could inform disruption activities. Finally, the fourth theme (community concerns) may be leveraged for the same ends by exploiting and amplifying heuristics such as the status quo bias, loss aversion bias and bounded rationality (Kahneman et al., 1991). Malicious AI innovations could be framed as risky or compromised, or stories of OPSEC failures caused by malicious AI could be planted in forum discussions.

## 6. Conclusion

Through our three main research questions and access to a diverse dataset, we clarified how cyber-offenders conceptualise AI tools and assess their relevance to their operational needs, which kinds of licit and illicit tools they have access to and favor, and what barriers prevent them from adopting AI more rapidly. Our findings support a more nuanced and complex narrative than the hyperbolic claims found in cybersecurity marketing material and sponsored media reports. The adoption of the diffusion of innovation framework allowed us to highlight how a small group of cybercrime innovators is re-inventing AI technology to scale their operations and make them more profitable, but also that the uncertainty inherent to the adoption of any innovation somehow interferes with quick, unrestrained adoption. It is still too early to predict how steep or flat the AI adoption curve will be in the cybercrime community, especially as its communication channels differ from those available to communities operating legally and the disruption activities of retardant agents could influence incentive and disincentive structures. Early evidence from the cryptocurrency intelligence firm Chainalysis suggests that the use of AI tools significantly boosts the scale and profitability of online scams (Chainalysis, 2026). Still, more data are needed to determine whether these results reflect a radical shift in cybercrime operations or a more limited pattern circumscribed to a small group of criminal innovators and early adopters. This is an important distinction, time playing a central role in diffusion processes, particularly in the case of AI, where the race to develop and commercialise new capacities—such as agentic AI—is fueled by hundreds of billions of dollars in investments, quickly making yesterday's innovation seem stale. From a cybercrime research perspective, this implies that data collection efforts cannot be confined to narrow time windows. Instead, longitudinal methodologies are needed to track how criminal innovations diffuse across adopter categories and to identify which policy and operational options are available and effective.

Sleegers, W., and Elsey, J. (2025), *Adoption and uses of LLMs among U.S. tech workers*, Rethink Priorities, San Francisco.

Sutherland, E. H. (1947), *Principles of criminology (Revised) Fourth edition*, J. B. Lippincott Co., Chicago.

Umbach, R., Henry, N., Beard, G. F., Berryessa, C. (2024), Non-consensual synthetic intimate imagery: Prevalence, attitudes, and knowledge in 10 countries, *CHI '24: Proceedings of the 2024 CHI Conference on Human Factors in Computing Systems*, 779, 1-20, doi: 10.1145/3613904.3642382.

Valeros, V., Sirokova, A., Catania, C., and Garcia, S. (2024), Towards better understanding of cybercrime: The role of fine-tuned LLMs in translation, *2024 IEEE European Symposium on Security and Privacy Workshops*, 8-12 July, Vienna, doi: 10.1109/EuroSPW61312.2024.00017.

Wall, D. (2015), Dis-organised crime: Towards a distributed model of the organization of cybercrime, *The European Review of Organized Crime*, *2*(2), 71-90, doi: 10.2139/ssrn.2677113.

Whelan, C., Jones, C, Martin, J., and Bright, D. (2025), The international structure of ransomware criminal groups: An analysis of organisational units, functions, roles and communication patterns within Conti, *Global Crime*, 1-23, doi: 10.1080/17440572.2025.2534387.

Whelan, C., Martin, J., Dupont, B., and Harkin, J. (2026), 'The name of the game': Policing perspectives on cybercrime disruption, *Police Practice and Research*, *27*(2), 289-306, doi: 10.1080/15614263.2025.2523522.

Wei, A., Haghtalab, N., and Steinhardt, J. (2023), Jailbroken: How does LLM safety training fail?, *37th Conference on Neural Information Processing Systems (NeurIPS 2023)*, 1-32, doi: 10.48550/arXiv.2307.02483.

Yampolskiy, R. (2024), *AI: Unexplainable, unpredictable, uncontrollable*, CRC Press, Boca Raton.

Yu, P. (2022), Diffusion of innovation theory, in F. Rapport, R. Clay-Williams, and J. Braithwaite (eds.), *Implementation science: The key concepts*, Routledge, London, 59-61, doi: 10.4324/9781003109945.

Yudkowsky, E., and Soares, N. (2025), *If anyone builds it, everyone will die*, Little, Brown and Company, Boston.
30

Figure 1. Child-code frequency and associated parent-theme frequency

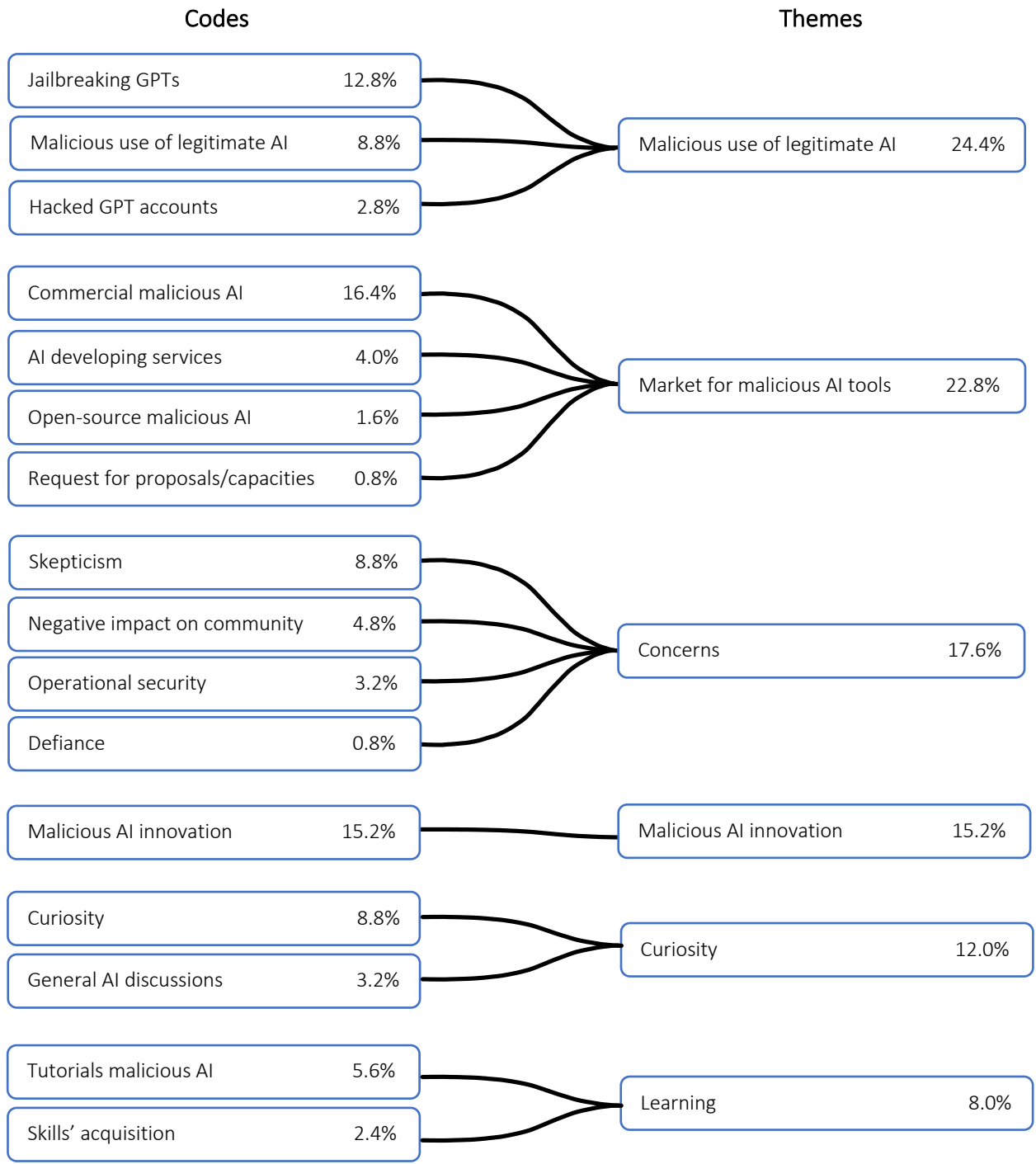

Figure 2. Stage of malicious AI diffusion based on adopter categories and innovativeness

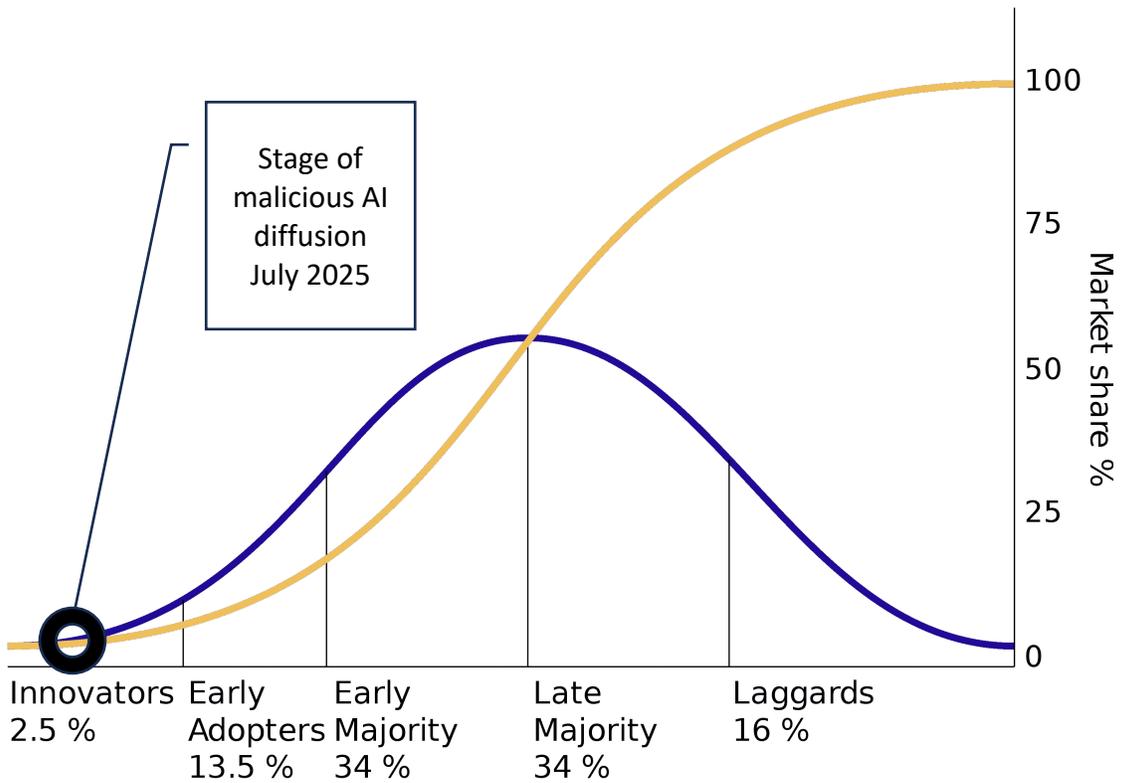

Source: Wikimedia Commons, based on Rogers, E. (1962), *Diffusion of innovations*, Free Press, London. Available from https://commons.wikimedia.org/wiki/File:Diffusion_of_ideas.svg.